\documentclass{pasj00}

\usepackage{times}

\begin{document}
\SetRunningHead{Totani}{FRBs from NS-NS mergers}
\Received{2013/07/18}
\Accepted{2013/08/16}

\title{Cosmological Fast Radio Bursts from Binary Neutron Star Mergers}

\author{Tomonori \textsc{Totani}\altaffilmark{1,2} }
\altaffiltext{1}{Department of Astronomy, School of Science,
  The University of Tokyo, Hongo, Bunkyo-ku, Tokyo 113-0033}
\altaffiltext{2}{Research Center for the Early Universe, 
  School of Science, The University of Tokyo, Hongo, Bunkyo-ku, Tokyo
  113-0033}


\maketitle

\begin{abstract}
  Fast radio bursts (FRBs) at cosmological distances have recently
  been discovered, whose duration is about milliseconds.  We argue
  that the observed short duration is difficult to explain by giant
  flares of soft gamma-ray repeaters, though their event rate and
  energetics are consistent with FRBs.  Here we discuss binary neutron
  star (NS-NS) mergers as a possible origin of FRBs.  The FRB rate is
  within the plausible range of NS-NS merger rate and its cosmological
  evolution, while a large fraction of NS-NS mergers must produce
  observable FRBs. A likely radiation mechanism is coherent radio
  emission like radio pulsars, by magnetic braking when magnetic
  fields of neutron stars are synchronized to binary rotation at the
  time of coalescence. Magnetic fields of the standard strength ($\sim
  10^{12-13}$ G) can explain the observed FRB fluxes, if the
  conversion efficiency from magnetic braking energy loss to radio
  emission is similar to that of isolated radio pulsars. Corresponding
  gamma-ray emission is difficult to detect by current or past
  gamma-ray burst satellites.  Since FRBs tell us the exact time of
  mergers, a correlated search would significantly improve the
  effective sensitivity of gravitational wave detectors.
\end{abstract}

\section{Introduction}

Thornton et al. (2013) reported a discovery of four enigmatic radio
transients emitting Jansky level flux during a few milliseconds,
called fast radio bursts (FRBs) (see also Lorimer et al. 2007; Keane
et al. 2012 for earlier events of possibly the same
population).  No repeated events were found, implying cataclysmic
nature.  The dispersion measure and the scattering signature seen in
the exponential tail of their light curves indicate that they are
located at cosmological distances of redshift $z \sim$ 0.5--1. No
counterparts were found at other wavelengths, and their origin is an
intriguing mystery.

The short duration of $\lesssim$ a few msec would place a strong
constraint on possible theoretical scenarios. Giant flares of soft
gamma-ray repeaters (SGRs) are suggested as a promising candidate by
Popov \& Postnov (2007) and Thornton et al. (2013), because of the
event rate consistent with that of FRBs (Ofek 2007) and sufficient
total energy to explain the FRB radio flux (Hurley et al. 2005; Palmer
et al. 2005; Terasawa et al. 2005).  However, radio emission only
within a msec scale duration is not naturally expected, though the
dynamical time of a neutron star is about 1 msec. Typical rotation
periods of SGRs are much longer ($\gtrsim$ 1 sec) and a continued
energy production at a giant flare would result in a longer radio
emission than the dynamical time.  Indeed, unsaturated gamma-ray light
curve of the 2005 giant flare of SGR 1806$-$20 shows a peak width of
$\sim$ 100 msec and subsequent modulation of the flux, indicating
repeated energy injections on a time scale longer than 100 msec
(Terasawa et al. 2005).

Other proposed scenarios for FRBs include interaction of supernova
shock and neutron star magnetosphere (Egorov \& Postnov 2009),
annihilating black holes (Keane et al. 2012), collapses of
supermassive rotating neutron stars (Falcke \& Rezzolla 2013), and
binary white dwarf mergers (Kashiyama et al. 2013).  In this letter we
discuss whether binary neutron star (NS-NS) mergers can explain the
new results about FRBs, with a different picture than those adopted in
previous studies about coherent radio emission from NS-NS mergers
(Lipunov \& Panchenko 1996; Hansen \& Lutikov 2001; Pshirkov \&
Postnov 2010; Lutikov 2013).

\section{Short Duration Radio Emission from Binary 
Neutron Star Mergers}

Electromagnetic (EM) wave signatures from NS-NS mergers have been
widely investigated in the literature (e.g., Metzger \& Berger 2012).
Two popularly discussed emission mechanisms, i.e., radioactivity of
ejected material (Li \& Paczy\'nski 1998; Metzger et al. 2010; Roberts
et al. 2011) and afterglows by energetic outflow interacting with
surrounding medium (Nakar \& Piran 2011; Shibata et al. 2011; Piran et
al. 2013), predict much longer time scales than msec and hence they
are not relevant to FRBs. Rather, coherent radio emission from neutron
star magnetosphere like isolated radio pulsars seems more plausible.
Hansen \& Lyutikov (2001) and Lyutikov (2013) considered pre-merger
coherent emission from interacting magnetosphere of a binary
consisting of a recycled, weak magnetic field neutron star and a
slowly rotating, strong field one, before their rotations are
synchronized to the binary period (see also Lipunov \& Panchenko 1996).
The radio flux predicted by Hansen \& Lyutikov (2001) is much lower
than those of FRBs, even if a very strong magnetic field ($10^{15}$ G)
is assumed. The flux estimated by Lyutikov (2013) is closer to the FRB
flux, but it seems difficult to reconcile the predicted time evolution
$\propto (-t)^{-1/4}$ with the observed short duration, where $-t$ is
the time before a merger whose minimum is $\sim$ msec.

Theoretically it is expected that rotations of neutron stars are not
tidally locked to the binary period until the last stage of merger
when tidal disruption starts (Bildsten \& Cutler 1992).  Therefore
strong coherent emission by rotation of individual neutron stars in a
binary is not expected during gravitational inspiral. However, at some
point in the last stage of a merger, their magnetic field
configuration should be synchronized with the binary rotation, and
coherent radio emission is expected by magnetic braking of a
misaligned rotating magnetic dipole, or plasma effect in the
magnetosphere, in a similar way to isolated radio pulsars.  If the
synchronization occurs before a completely merged object forms, the
radiation would come from the original magnetic dipole of a neutron
star. Instead, a magnetic field of a merged object may be responsible
for the radio emission, and in this case magnetic field strength may
be amplified by differential rotation (Pshirkov \& Postnov 2010;
Shibata et al. 2011).

Because of the strong dependence of the magnetic breaking energy loss
rate on rotation period ($\dot{E} \propto P^{-4}$), the luminosity
should sharply increase with the synchronization of magnetic fields,
and such coherent emission will continue until the merged neutron
stars form a black hole (BH). Thus, this radiation mechanism seems
favorable to explain the msec scale duration of FRBs. The expected
energy loss rate by magnetic braking can be estimated from the
standard magnetic dipole radiation formula, using typical magnetic
field strength and radius of a neutron star, but with binary orbital
period at the time of coalescence, i.e., about milliseconds:
\begin{eqnarray}
\dot E &=& - 6.2 \times 10^{45}  
\left( \frac{B}{10^{12.5} \rm \ G} \right)^{2} \nonumber
\left( \frac{R}{10 \rm \ km} \right)^{6} \\
&& \times \left( \frac{P}{0.5 \rm \ msec} \right)^{-4} \ \rm erg \ s^{-1} \ .
\end{eqnarray}

Pshirkov \& Postnov (2010) also considered coherent emission by
magnetic breaking energy loss, but they considered a strongly
amplified magnetic field ($\sim 10^{15}$ G) to explain the luminosity
($\sim 10^{50-52}$ erg/s) of short gamma-ray bursts (GRBs) that are
much brighter but rarer than FRBs.  We show below that FRBs can be
explained without strong $B$ field amplification, though a significant
fraction of NS-NS merger events should produce FRBs.

\section{Rate and Expected Radio Flux}

The rate of FRBs is estimated to be $1.0^{+0.6}_{-0.5} \times 10^4 \
\rm day^{-1} sky^{-1}$ (Thornton et al. 2013), which is translated
into a rate per unit comoving volume of $2.3^{+1.4}_{-1.2} \times 10^4
\ \rm yr^{-1} Gpc^{-3}$ to the comoving distance $D_{\rm comv}$ = 3.3
Gpc corresponding to maximum redshift of $z_{\max}=1$.  This rate is
statistically consistent with the ``plausible optimistic estimate'' of
NS-NS mergers, $\sim 10^4 \ \rm yr^{-1} Gpc^{-3}$, as reviewed by
Abadie et al. (2010).  The FRB rate estimated from the four events is
still statistically highly uncertain.  Choosing a slightly larger
value of $z_{\max}$ will further reduce the rate in proportion to
$D_{\rm comv}^{-3}$.  It should be noted that this comparison does not
take into account the cosmological effects (the cosmological time
dilation and merger rate evolution).  Increase of the merger rate by a
factor of $\sim 4$ from $z$ = 0 to 1 is reasonable from the cosmic
star formation history (Totani 1997; Dominik et al. 2013), which is a
bigger effect than the cosmological time dilation that reduces the
observed rate by $\propto (1+z)^{-1}$, making the FRB rate closer to
the standard ``realistic'' estimate of NS-NS mergers ($10^3 \ \rm
yr^{-1} Gpc^{-3}$).  Therefore the observed FRB rate is consistent
with the NS-NS merger scenario, but the fraction of NS-NS mergers
producing observable FRBs must be of order unity.

The coherent radio emission discussed above may be observed from most
of the merger events, if the emission is not strongly beamed.  A
beaming fraction $\Omega / (4 \pi)$ of order unity is not unreasonable
given the theoretical uncertainty about beaming of pulsar
radio emission (e.g., Kalogera et al. 2001), where $\Omega$ is the
total opening solid angle of radio emission.  A direct, purely
observational estimate based on statistics of associations between
pulsars and pulsar-powered nebulae indicates a beaming fraction of
$\sim$ 60\% for young pulsars (Frail \& Moffett 1993).

Since the physics of coherent radio emission from pulsars is poorly
understood (Lyubarsky 2008), we discuss the expected radio flux in
terms of the ratio $\epsilon_r$ of the radio luminosity $\nu L_\nu$ to
the total energy loss rate $|\dot E|$.  Although there is a large
scatter of $\epsilon_r$ for isolated radio pulsars in our Galaxy, we
choose a typical value of $\epsilon_r = 10^{-4}$ (Taylor et al. 1993;
Manchester et al. 2005) at the rest-frame frequency corresponding to
the observed frequency $\nu_{\rm obs}$ = 1.4 GHz.  Using a luminosity
distance $D_{\rm lum}$ to $z = 0.75$, we get
\begin{eqnarray}
  F_\nu &=& \frac{1}{\nu_{\rm obs}} 
  \frac{\epsilon_r |\dot{E}|}{4 \pi D_{\rm lum}^2} 
  = 0.02
  \left( \frac{\epsilon_r}{10^{-4}} \right) 
  \left( \frac{D_{\rm lum}}{4.6 \rm \ Gpc}
  \right)^{-2} \nonumber \\
  && \times
  \left( \frac{B}{10^{12.5} \ \rm G} \right)^{2}
  \left( \frac{R}{10 \ \rm km} \right)^{6}
  \left( \frac{P}{0.5 \ \rm msec} \right)^{-4} 
  \ \rm Jy \ .
\end{eqnarray}
Therefore, the observed FRB fluxes ($\sim$ 0.5 Jy) can be
explained by a slightly higher radio emission efficiency of
$\epsilon_r \sim 10^{-3}$, or a modest amplification of magnetic field
strength to $B \sim 10^{13}$ G, but a very strong magnetic field such
as $10^{15}$ G is not necessary if $\epsilon_r$ is similar to those
for isolated radio pulsars. If magnetic fields of both neutron stars
before a merger are much weaker than $10^{12}$ G as a result of
magnetic field decays, amplification in a merged object would be
required, though decay of magnetic fields in neutron stars is
still highly uncertain (e.g., Mukherjee \& Kembhavi 1997).

\section{Discussion}

Several predictions and implications can easily be derived from the
hypothesis proposed here.

The NS-NS merger rate must be close to the optimistic (but still
plausible) estimate, which is certainly a good news to gravitational
wave astronomy. The ``plausible optimistic estimate'' of NS-NS merger
rate in Abadie et al. (2010) is still one order of magnitude lower
than the current upper limit (Abadie et al. 2012), but such a high
rate predicts a detection within a few years in the early
commissioning phase of the Advanced LIGO, whose ultimate detection
rate would be $\sim 400 \ \rm yr^{-1}$ to $\sim$ 200 Mpc (Abadie et
al. 2010; Aasi et al. 2013).  Typical radio flux of FRBs at 200 Mpc
would be about 100 Jy, though a large scatter of radio flux from event
to event is expected by variation of $\epsilon_r$, as inferred from
radio pulsars.  An important advantage of FRBs compared with longer
time-scale EM signals of NS-NS mergers is that FRBs tell us the exact
time of coalescence. If a nearby/bright FRB sample is constructed by
future radio transient surveys, searching for gravitational wave
bursts correlated with FRBs would significantly improve the effective
sensitivity of gravitational wave detectors, allowing to detect more
distant mergers.  Another advantage of FRBs for gravitational wave
astronomy is that they are expected to be observable for most of NS-NS
merger events, in contrast to e.g., short GRBs.

If host galaxies are identified for FRBs, they should include early
type galaxies that are not star forming, while the SGR or supermassive
neutron star scenarios predict only star-forming galaxies. Since FRBs
can be observed to more distant universe than gravitational wave
bursts, the cosmological rate evolution of NS-NS mergers and its
relation to host galaxy evolution may be studied in the future, which
would be complementary to short GRBs (if they are also produced by
NS-NS mergers).

Detectability of FRBs in other wavelength is also of interest.  Pulsed
gamma-ray luminosity of pulsars is typically $\sim 10$\% of spin-down
luminosity (Abdo et al. 2010), and msec duration gamma-ray emission
from FRBs may be detected by GRB satellites.  Assuming
$\epsilon_\gamma/\epsilon_r = 10^3$ for the gamma-ray band, a typical
radio flux of 0.5 Jy at 1.4 GHz is corresponding to a gamma-ray flux
of $\nu F_\nu \sim 7 \times 10^{-12} \ \rm erg \ cm^{-2} s^{-1}$, which
is much fainter than the typical {\it Swift} trigger threshold of
$\sim 10^{-8} \ \rm erg \ cm^{-2} s^{-1}$ in 15--150 keV (Sakamoto et
al. 2011).  Note that the flux threshold for msec duration bursts
should be much higher than this (T. Sakamoto, a private
communication). The 50--300 keV flux threshold of BATSE GRBs in the 64
msec trigger window is $\sim 10^{-7} \rm \ erg \ cm^{-2} s^{-1}$
(Fishman et al. 1994), and FRBs of 1 msec duration must be very close
($\sim$ Mpc) to be detected by this trigger condition, but the
expected event rate in such a small volume is extremely small.  A
search optimized for msec duration bursts in the past GRB data may be
interesting.

If short GRBs are also produced by NS-NS mergers, their small rate
($\sim 10 \ \rm yr^{-1} Gpc^{-3}$, Coward et al. 2012) compared with
FRBs indicates that only a tiny fraction of NS-NS merger events
produce observable short GRBs.  This is possible if short GRBs are
strongly beamed, and/or they are produced by rarer events such as
large mass ratio binaries, mergers resulting in a hypermassive neutron
star (HMNS) supported by rotation, or NS-BH/BH-BH mergers (Shibata \&
Taniguchi 2006; Faber \& Rasio 2012).  The beaming-corrected estimates
of short GRB rate can be as high as $\sim 1000 \ \rm yr^{-1} Gpc^{-3}$
(Coward et al. 2012; Enrico Petrillo et al. 2013), and if this is
correct, about 10\% of NS-NS merger events are producing short GRBs to
a certain direction.  On the other hand, our hypothesis predicts that
most of short GRBs must be associated with FRBs. The time delay by
dispersion is about 1 second at GHz bands, but longer time delay at
lower frequencies may allow to detect FRBs by follow-up searches after
short GRBs (Lipunova et al. 1997; Pshirkov \& Postnov 2010).

Sub-msec scale fine temporal features are difficult to see in the four
events of Thornton et al. (2013), because of dispersive smearing in the
frequency-integrated flux and scattering tails produced by propagation
in intergalactic medium.  Future high time resolution observations of
bright/nearby FRBs may reveal quasi-oscillatory behavior, reflecting
binary orbital period or rotation rate of merged objects, though
numerical studies are necessary to make quantitative predictions.  The
observed short duration implies that merged neutron stars should
promptly form a black hole on a dynamical time scale. This picture is
consistent with the latest numerical simulations of NS-NS mergers, but
a fraction of merger events may form a HMNS that survives for a time
scale longer than milliseconds before collapsing into a black hole
(Hotokezaka et al. 2011; Faber \& Rasio 2012).  Such a HMNS may
radiate a pulsar-like periodic coherent emission during its lifetime,
and the fraction of such relatively long FRBs may give constraints on
the mass distribution of NS-NS binaries and the equation of state at
nuclear densities.

\bigskip

The author would like to thank Y. Itoh, K. Kashiyama, N. Kawanaka and
T. Sakamoto for useful comments.

\end{document}